\documentclass[epsfig,12pt]{article}
\usepackage{amsmath,amssymb}
\usepackage{epsfig}

\newcommand{\1}{{\sf 1 \!\! 1}}

\newcommand{\OM}{\vec s}
\newcommand{\aaa}{a}
\renewcommand{\tau}{t}
\newcommand{\mvec}[2]{\left(\begin{array}{c}#1\\ #2\end{array}\right)}

\newcommand{\mat}[4]{\left(%
\begin{array}{cc}#1&#2 \\#3&#4\end{array}\right)}

\newcommand{\bra}[1]{\left\langle#1\right|}
\newcommand{\ket}[1]{\left|#1\right\rangle}
\newcommand{\bracket}[2]{\left\langle\left.#1\right|#2\right\rangle}

\makeatletter
\@addtoreset{equation}{section}
\makeatother

\newcommand{\sfig}[4]{                                                  %
   \begin{figure} [tb] \begin{center} #1 \end{center}                   %
      \centerline{ \parbox{420pt}{\caption[#2]{{\sf #3}} \label{#4}}}   %
   \end{figure}}                                                        %

\title{Low-Energy Effective Theories of Quantum Spin and Quantum Link Models
\footnote{This work is supported in part by funds provided by the U.S.
Department of Energy (D.O.E.) under cooperative research agreement
DE-FC02-94ER40818.}}
\author{B. Schlittgen$^a$ and U.-J. Wiese$^b$
\\ \\
$^a$ Department of Mathematics,\\ Massachusetts Institute of Technology, \\
Cambridge, Massachusetts 02139 \\ \\
$^b$ Center for Theoretical Physics, Laboratory for Nuclear Science \\
and Department of Physics, Massachusetts Institute of Technology, \\
Cambridge, Massachusetts 02139}

\begin{document}
\maketitle
\begin{abstract} \normalsize

Quantum spin and quantum link models provide an unconventional regularization
of field theory in which classical fields arise via dimensional reduction of
discrete variables. This D-theory regularization leads to the same continuum 
theories as the conventional approach. We show this by deriving the low-energy
effective Lagrangians of D-theory models using coherent state path integral 
techniques. We illustrate our method for the $(2+1)$-d Heisenberg quantum spin
model which is the D-theory regularization of the 2-d $O(3)$ model. Similarly,
we prove that in the continuum limit a $(2+1)$-d quantum spin model with 
$SU(N)_L\times SU(N)_R\times U(1)_{L=R}$ symmetry is equivalent to the 2-d principal chiral model.
Finally, we show that $(4+1)$-d $SU(N)$ quantum link models reduce to ordinary
4-d Yang-Mills theory.

\end{abstract}
 
\maketitle

\newpage

\section{Introduction}

In the conventional approach to quantum field theory the fundamental degrees of
freedom are continuous classical fields. To regularize the ultraviolet 
divergences beyond perturbation theory it is natural to introduce a space-time 
lattice. For example, in Wilson's lattice gauge theory the gluons are 
represented by classical $SU(N)$ parallel transporter matrices. Similarly, the
fundamental degrees of freedom of a lattice $O(3)$ model are classical unit 
vectors. In this paper we use an alternative approach to field theory ---
D-theory --- in which classical fields are replaced by discrete 
variables (quantum spins or quantum links) that undergo dimensional reduction
\cite{ChWi97,BBCW98,BChW97,Wies98}.

For example, at zero temperature the $O(3)$ symmetries of both, the $(2+1)$-d 
ferromagnetic and antiferromagnetic Heisenberg quantum spin models break 
spontaneously, giving rise to massless Goldstone bosons --- the so-called 
magnons or spin waves. These 
magnons are collective excitations of many quantum spins and are effectively
described by a continuous classical field. It is remarkable that these 
continuous degrees of freedom emerge from a microscopic theory of purely 
discrete quantum spins. The low-energy effective theory of magnons is an $O(3)$
model in $(2+1)$ dimensions. At small, non-zero temperature, and hence at 
finite extent 
$\beta$ of the Euclidean time dimension, the correlation length $\xi$ of the 
Goldstone bosons is large compared to $\beta$ and hence the system undergoes
dimensional reduction to the 2-d $O(3)$ model. In this case, the 
Coleman-Hohenberg-Mermin-Wagner theorem \cite{MeWa66} implies that $\xi$ must 
become 
finite and that the magnons pick up a nonperturbatively generated mass gap 
$m = 1/\xi$. As a consequence of asymptotic freedom of the 2-d $O(3)$ model, 
$\xi$ is exponentially large in $\beta$, $\xi \sim \exp(2 \pi \rho_s \beta)$, 
where $\rho_s$ is the spin stiffness of the underlying quantum spin system
\cite{CHaN88,Has91,KoCh89}. Hence,
$\rho_s \beta = 1/g^2$ plays the role of the coupling constant of the 
dimensionally reduced theory. The continuum limit of that theory is reached by
varying the extent $\beta$ of the extra dimension, not by adjusting a bare
coupling constant.

Dimensional reduction of discrete variables is not limited to the quantum
Heisenberg model. In fact, it is a generic phenomenon that gives rise to the
D-theory formulation of field theory. For example, as we show in this paper,
the 2-d principal chiral model that is traditionally formulated in terms of 
continuous classical $U(N)$ matrix fields can also be expressed as a system of 
generalized quantum spins in $(2+1)$ dimensions (cf.\ \cite{ScWi00}). The 
components of these 
quantum spins are generators of an $SU(2N)$ algebra. Again, the discrete 
variables undergo dimensional reduction to 2-d if the $(2+1)$-d system has 
massless Goldstone bosons. We identify appropriate representations of the
$SU(2N)$ algebra for which this is indeed the case.

Gauge theories can be formulated in terms of discrete quantum links which are
gauge covariant generalizations of quantum spins. A quantum link is an
$N \times N$ parallel transporter matrix whose elements are generators of 
$SU(2N)$. The dimensional reduction of quantum link models works differently
from the case of quantum spins. While in the spin case an infinite correlation 
length arises as a result of the spontaneous breakdown of a global symmetry, 
for gauge theories spontaneous symmetry breaking leads to a massive Higgs 
phase with a finite correlation length. Moreover, confined phases in 
non-Abelian gauge theories also have finite correlation lengths and therefore 
do not lead to dimensional reduction. However, gauge theories in five 
dimensions can exist in a non-Abelian Coulomb phase with massless unconfined
gluons \cite{BBCW98,Creu79}. The massless gluons of a $(4+1)$-d quantum link 
model are collective
excitations of many quantum links, just as magnons are collective excitations
of quantum spins. If a $(4+1)$-d quantum link model exists in a Coulomb phase 
for an infinite extent $\beta$ of the fifth dimension, it will undergo 
dimensional reduction to an ordinary 4-d gauge theory once $\beta$ becomes 
finite \cite{ChWi97}. This is a consequence of the confinement hypothesis, 
which is the gauge analog of the Coleman-Hohenberg-Mermin-Wagner theorem of 
the spin 
case. In particular, a gluon cannot remain massless when $\beta$ becomes 
finite because it then effectively lives in a 4-d world and hence should be 
confined. The corresponding finite correlation length $\xi$ is related to the 
glueball mass $m = 1/\xi$. As before, asymptotic freedom of 4-d non-Abelian 
gauge theories implies that $\xi$ is exponentially large in $\beta$, $\xi \sim 
\exp(8 \pi^2 \beta/11 e^2 N)$, where $e$ is the dimensionful gauge coupling of 
the underlying $(4+1)$-d quantum link model. Hence, the role of the coupling
constant of the dimensionally reduced theory is played by $\beta/e^2 = 1/g^2$.
So again, the continuum limit of the theory is reached by varying the extent 
$\beta$ of the extra dimension, not by adjusting a bare coupling constant.

As just explained, taking the continuum limit of a quantum link model requires
a fifth dimension in order to obtain a large gluonic correlation length. It is
then very natural to make use of the fifth dimension to include quarks as
domain wall fermions. In particular, Shamir's variant \cite{Sham93,FuSh95}
of Kaplan's original proposal \cite{Kapl92} provides a suitable realization 
of full quantum link QCD. Again, the
correlation length of the quarks is controlled by the extent of the fifth
dimension and is exponentially large in $\beta$. Consequently, one reaches both
the chiral and the continuum limit by sending $\beta$ to infinity. This 
requires no fine-tuning of bare coupling constants, which makes D-theory an 
attractive alternative to the traditional approach to field theory
\cite{BChW97}.

Models with discrete degrees of freedom have been studied before.
Generalizations of antiferromagnetic quantum spin models were discussed
by Read and Sachdev in \cite{ReSa89} for the case of an $SU(N)$ symmetry 
group. In \cite{DaSS93}, Radjbar-Daemi, Salam and Strathdee considered 
discrete spin systems with a general
symmetry group and showed how the continuum limit of such theories
corresponds to sigma-model-type field theories. They considered both
ferro- and antiferromagnetic cases. In a follow up to that paper, they also
investigated the renormalization group flow of a particular continuum theory
on the manifold $SU(3)/[U(1)\times U(1)]$ \cite{DaSt93}. 
A quantum link model with a $U(1)$ gauge symmetry was first constructed by 
Horn in \cite{Horn81}. Orland and Rohrlich 
introduced an $SU(2)$ quantum link model \cite{OrRo90}.
In the present context, it was realized in \cite{ChWi97} that models with 
discrete variables
can give rise to ordinary field theories, including QCD \cite{BChW97}, via 
dimensional reduction. A detailed analysis of how the physics of conventional
lattice gauge theory with $U(1)$ gauge group is reproduced by the
$U(1)$ quantum link model is given in \cite{Chan99}.

A key issue in the previous discussion is the existence of massless Goldstone
bosons in quantum spin models, and of massless Coulombic gauge bosons in 
quantum link models. In the $(2+1)$-d spin $1/2$ antiferromagnetic quantum 
Heisenberg model it was unclear for some time if the $O(3)$ symmetry is 
spontaneously broken. For
larger spin representations, however, one can prove analytically that this is 
indeed the case. By now, detailed numerical simulations have shown that 
spontaneous symmetry breaking also occurs for spin $1/2$ \cite{Barn91,WiYi94}. 
For the $(2+1)$-d $SU(N)_L\times SU(N)_R\times U(1)_{L=R}$ symmetric
quantum spin model constructed in this 
paper it is a priori not clear if spontaneous symmetry breaking occurs, and
thus if massless Goldstone bosons exist. Here we show that for a
sufficiently large representation of the embedding algebra $SU(2N)$ with a
rectangular Young tableau, this model becomes a Wilson-type lattice 
principal chiral model in 3-d. Such a model is known to exist in a phase
of spontaneously broken symmetry \cite{Dreh92}
and hence we conclude that at finite
extent $\beta$ of the third dimension it
undergoes dimensional reduction to the usual 2-d 
$SU(N)_L\times SU(N)_R\times U(1)_{L=R}$ principal chiral model. Similarly, 
we show that for a sufficiently large representation of 
$SU(2N)$ the low-energy behavior of a $(4+1)$-d quantum link model is that
of a Wilson-type lattice gauge theory in 5 dimensions. 
From numerical simulations, we know that such a model can exist 
in a non-Abelian Coulomb phase \cite{BBCW98,Creu79} and
hence undergoes dimensional reduction to a 4-d $SU(N)$ Yang-Mills theory
once the extent of the fifth dimension becomes finite. This
proves that D-theory is indeed a valid regularization of these models. It would
be interesting and of practical importance to investigate if the massless 
phases arise also for small representations of $SU(2N)$. This requires detailed
future numerical studies.

The rest of this paper is organized as follows. In chapter \ref{ch:O(3)} we use
the quantum Heisenberg model to illustrate D-theory with a simple example. 
Chapter \ref{ch:U(N)} contains the construction of quantum spin and
quantum link models with $SU(N)_L\times SU(N)_R\times U(1)_{L=R}$ symmetry 
embedded in the algebra 
of $SU(2N)$. In chapter \ref{ch:coherent} a coherent state formalism is
presented in order to describe the large representation limit of $SU(2N)$. The
existence of massless Goldstone bosons for the 
$SU(N)_L\times SU(N)_R\times U(1)_{L=R}$ symmetric quantum spin
model is shown in chapter \ref{ch:principal} and the existence of a massless
Coulomb phase for $SU(N)$ quantum link models is derived in chapter 
\ref{ch:gauge}. Finally, chapter \ref{ch:conclusions} contains our conclusions.

Throughout the paper we distinguish classical degrees
of freedom, denoted by lower case letters, from quantum operators, denoted
by upper case letters.

\section{From the $(2+1)$-d Heisenberg Quantum Spin Model to the 2-d $O(3)$
Model}
\label{ch:O(3)}

To motivate the ideas leading to the D-theory formulation of field 
theory, let us review the well-known relation between the $(2+1)$-d Heisenberg
quantum spin model and the 2-d $O(3)$ model. The 2-d $O(3)$ model is 
asymptotically free, it has a nonperturbatively generated mass gap, as well as 
instantons and $\theta$-vacua and has been used as a simple toy model for QCD 
in four dimensions. 

The continuum action of the 2-d $O(3)$ model is given by
\begin{equation}
S[\vec s\,] = \frac{1}{2g^2} \int d^2 x \, 
\partial_\mu \vec s \cdot \partial_\mu \vec s,
\end{equation}
where $\vec s$ is a classical, 3-component unit vector that assumes 
continuous values. Clearly, this action is invariant under global $O(3)$ 
transformations $\vec s\,' =  R \vec s$, where $R^T R = \1$. The standard 
procedure to regularize this theory beyond perturbation theory is to follow
Wilson and introduce a lattice as an ultraviolet cut-off. Partial derivatives 
are then replaced by finite differences and (after dropping an irrelevant 
constant), the lattice action takes the form,
\begin{equation}
S[\vec s\,]= - \frac{1}{g^2} \sum_{x,\mu} \vec s_x \cdot \vec s_{x+\hat\mu}.
\end{equation}
Here, $\hat\mu$ is the unit-vector in the $\mu$-direction.
The theory is quantized by writing down the partition function, which is a path
integral over classical field configurations,
\begin{equation}
Z = \int {\cal D}\vec s \,\exp\left(- S[\vec s\,]\right).
\end{equation}
Due to asymptotic freedom, the continuum limit of the lattice-regularized 
theory is attained by taking the bare coupling constant $g$ to zero. In this 
limit, the correlation length $\xi \sim \exp(2 \pi/g^2)$ diverges 
exponentially, thus eclipsing any short-distance lattice artifacts.

D-theory follows a radically different approach to field quantization. Instead
of performing a path integral over continuous classical fields, those fields
are replaced by discrete quantum variables. For example, the above 2-d $O(3)$
field theory is formulated in terms of quantum spins with a Heisenberg model
Hamiltonian
\begin{equation}
H = J \sum_{x,\hat\mu} \vec S_x \cdot \vec S_{x+\hat\mu}.
\end{equation}
The components of the spin vectors $\vec S$ are the generators of $SO(3)$ and 
they satisfy the usual commutation relations 
\begin{equation}
[S_x^i,S_y^j] = i \delta_{xy} \varepsilon_{ijk} S_x^k.
\end{equation}
Notice that we are free to choose any representation of $SO(3)$ for the 
generators $\vec S_x$, 
not just spin $1/2$. The $SO(3)$ symmetry of the quantum 
Hamiltonian is expressed as $[H,\vec S] = 0$, where $\vec S = \sum_x \vec S_x$.

The case $J < 0$ corresponds to a ferromagnet and $J > 0$ to an antiferromagnet
with a N\'eel-ordered ground state. We restrict our attention to the former
case. The partition function for the Heisenberg model is given by
\begin{equation}
\label{eq:part_fun}
Z= \text{Tr} \exp(- \beta H),
\end{equation}
where the trace is taken in a large Hilbert space, the direct product of the 
Hilbert spaces corresponding to individual lattice sites. The Hamiltonian 
evolves the system in an extra dimension, giving rise to a $(2+1)$-dimensional 
field theory. For a condensed matter quantum spin system the additional
dimension is Euclidean time. In D-theory, however, Euclidean time is part of
the 2-d lattice and the additional Euclidean dimension will ultimately
disappear via dimensional reduction. 

As discussed in the introduction, dimensional reduction requires an infinite
correlation length, which in this case is due to the existence of massless 
Goldstone bosons. One way of addressing the question of symmetry breaking
and hence dimensional reduction, is to investigate the limit of large spin $S$.
To do this, we set up a spin coherent state representation of the path 
integral as discussed in \cite{Auer94}. 
The highest weight vector of the representation with spin $S$ is 
denoted by $|S,S\rangle$. 
To generate a system of coherent states, we must 
therefore act with all group 
elements on this state. A general $SO(3)$ group element 
can be parametrized
using the three Euler angles as
\begin{equation}
R(\chi,\theta,\phi) = \exp(i \phi S^3) \exp(i \theta S^2)
\exp(i \chi S^3).
\end{equation}
Hence, we obtain the system of coherent states, with the
following parametrization,
\begin{equation}
\ket{\OM\,} = R(\chi,\theta,\phi) \ket{S,S}
= \exp(i \phi S^3) \exp(i \theta S^2) \exp(i \chi S^3) 
\ket{S,S}.
\end{equation}
Notice that $e^{i\chi S^3}\ket{S,S} =e^{i\chi S}
\ket{S,S}$, generating the isotropy subgroup for the
highest weight vector, $\ket{S,S}$. Choosing
$\chi\equiv 0$ then corresponds to taking a cross section
in the fiber bundle with base $X=SO(3)/SO(2)$ and fiber
$SO(2)$.

The coherent states are now parametrized only by $\theta$
and $\phi$, which fall in the ranges $\theta\in
[0,\pi]$ and $\phi\in [-\pi, \pi)$, so that we can think
of $\OM$ as parametrizing a vector on the unit sphere,
$\OM=(\sin\theta\cos\phi, \sin\theta\sin\phi,
\cos\theta)$.
To obtain an expression for the coherent states as a 
superposition of $S^3$-eigenstates, we introduce
Schwinger bosons ---
$a,a^{\dag}$ and $b,b^{\dag}$ --- which are 
two sets of boson creation and annihilation operators,
satisfying the usual commutation relations.
We can then write,
\begin{equation}
S^1 +iS^2=a^{\dag}b,\;\;\;
S^1 -iS^2=b^{\dag}a,\;\;\;
S^3=\frac{1}{2}(a^{\dag}a-b^{\dag}b).
\end{equation}
The additional constraint $n_a +n_b=2S$ fixes the 
representation of spin $S$. We can 
use the raising and lowering operators to generate
the other weight vectors,
\begin{equation}
\ket{S,m} = \frac{(a^{\dag})^{S+m}}{\sqrt{(S+m)!}}
\frac{(b^{\dag})^{S-m}}{\sqrt{(S-m)!}}
\ket{0} .
\end{equation}  
To rotate such a state by $R$ as above, we note that
\begin{align}
\mvec{a^{\dag}}{b^{\dag}}'&=
\mvec{R\, a^{\dag}\, R^{-1}}{R \,b^{\dag}\, R^{-1}} 
=\exp\left(i\mbox{$\frac{\chi}{2}$}\,\sigma^z\right)
   \exp\left(i\mbox{$\frac{\theta}{2}$}\,\sigma^y\right)
   \exp\left(i\mbox{$\frac{\phi}{2}$}\,\sigma^z\right)
   \mvec{a^{\dag}}{b^{\dag}} \notag \\
&=\mat{u \exp\left(i\frac{\chi}{2}\right)}
 {v    \exp\left( i\frac{\chi}{2}\right)}
{-v^{*}\exp\left(-i\frac{\chi}{2}\right)}
 {u^{*}\exp\left(-i\frac{\chi}{2}\right)}\mvec{a^{\dag}}{b^{\dag}}.
\end{align} 
Here, $u(\theta,\phi)=\cos(\theta/2)\exp\left(i\phi/2\right)$
and $v(\theta,\phi)=\sin(\theta/2)\exp\left(-i\phi/2\right)$.
This leads to the following expression for the coherent
states (recall that $\chi\equiv 0$),
\begin{equation}
\label{eq:cs_vec}
\ket{\OM\,} =\frac{(a^{\dag'})^{2S}}{\sqrt{(2S)!}}
\ket{0} =\sqrt{(2S)!}\sum_m
\frac{u^{S+m}v^{S-m}}{\sqrt{(S+m)!(S-m)!}}\ket{S,m}.
\end{equation}
These coherent states are not all linearly independent
--- they form an over-complete set of states. In
particular, we can express the identity operator
as a superposition of coherent states. The measure
of integration we use here is $[(2S+1)/4\pi]d\OM =[(2S+1)/4\pi]
\sin\theta \,d\theta\, d\phi$,
\begin{equation}
\label{eq:res_id}
\frac{2S+1}{4\pi}\int d\OM \ket{\OM\,}\bra{\OM\,}=
\sum_m \ket{S,m}\bra{S,m}=\1.
\end{equation}
Another important property is the following,
\begin{equation}
\label{eq:cs_prop1}
\OM\cdot \vec S\ket{\OM\,}=S\ket{\OM\,}.
\end{equation}

A system of coherent states in the
large Hilbert space on which the entire Hamiltonian acts
is simply given by a direct
product of the coherent state systems we have derived
for each lattice site.
From property (\ref{eq:cs_prop1}) 
it is straightforward to obtain
an expression for the expectation value
of the Heisenberg Hamiltonian in a coherent state,
\begin{equation}
\label{eq:ham_expec}
{\cal H}[\OM\,]=\bra{\OM\,}H \ket{\OM\,}=
\frac{S^2 J}{2}\sum_{x,\hat{\mu}} \OM_x\cdot
\OM_{x+\hat{\mu}}.
\end{equation}

Using these ingredients, we can express the 
partition function (\ref{eq:part_fun}) as a path integral
over coherent states. The standard procedure is to divide
up the ``time'' interval $\beta$ into $N_{\varepsilon}$ 
small intervals of width $\varepsilon=\beta/N_{\varepsilon}$,
and to insert a resolution of the identity 
(\ref{eq:res_id}) in between each time slice.
Eventually, we take $N_{\varepsilon}\to\infty$.
We can manipulate the expression for the path integral using  
(\ref{eq:ham_expec}).
Also, we write
\begin{equation}
\frac{\OM_x(\tau+\varepsilon)-\OM_x(\tau)}{\varepsilon}
\to \dot{\OM}_x +{\cal O}(\varepsilon).
\end{equation}
In our parametrization of the coherent states, the overlap between 
neighboring states is given by
\begin{equation}
\bracket{\OM(\tau+\varepsilon)}{\OM(\tau)}=
\exp\left( -iS\varepsilon\sum_x  \dot{\phi}_x
\cos(\theta_x)\right).
\end{equation}
We thus get the path integral
\begin{equation}
Z=\int{\cal D}\OM \,\exp\left(-S[\OM\,]
\right),
\end{equation}
where
\begin{equation}
\label{eq:ber}
S[\OM\,]
=iS\sum_x \omega[\OM_x]+\int_0^{\beta}d\tau \,
{\cal H}[\OM\,],
\end{equation}
and $\omega[\OM\,]=\int_0^{\beta}d\tau\, \dot{\phi}
\cos\theta
=\oint_{\phi_0}^{\phi_0}d\phi\, \cos(\theta_{\phi})$
is a Berry phase term. The geometric nature of the Berry phase term
is evident, as it depends only on the path on the unit sphere traced out by
the spin, and not on the explicit dependence of this path on $\tau$. 
In fact, this term measures the area enclosed by the path $\vec s(\tau)$ 
on $S^2$. It may be written in gauge invariant form as
\begin{equation}
\omega[\OM\,]=\int_0^{\beta}d\tau \int_0^1 dv\, \vec s\cdot 
\left(\partial_{\tau}\vec s \times \partial_v \vec s\,\right).
\end{equation}
We have introduced the interpolating field $\vec s(\tau,v)$, which depends on 
an additional (fourth) dimension and obeys the boundary conditions 
$\vec s(\tau,1)=\vec s(\tau)$, $\vec s(\tau,0)=\vec s(\tau',0)$ 
and $\vec s(0,v)
=\vec s(\beta,v)$. This field is therefore parametrized on a disc with $v=1$
as the boundary. The only requirement on the interpolation 
$\vec s(t,v)$ from the boundary of the disc to the interior is that it be 
a smooth function of $v$. In particular, 
the value of the Berry phase term differs
from one chosen interpolation to another only by an integer multiple of
$4\pi$. This is a direct consequence of the fact that the second homotopy 
group of $S^2$ is $\Pi_2(S^2)=\mathbb{Z}$. Hence, for the value of the path 
integral to be independent of the chosen interpolation,
the spin $S$ needs to be quantized in half-integer units.
With the simple parametrization $s^i(\tau,v)=vs^i(\tau)$ for 
$i=1,2$, we can perform the integral over $v$ to obtain an expression for
the  Berry phase
that is equivalent to the one given in (\ref{eq:ber}), and hence arrive
at the following continuum action,
\begin{equation}
\label{eq:O(3)_eff_ac}
S[\OM\,] = \int_0^{\beta}d\tau\int d^2 x\,
\left[iS(1+s^3)^{-1}(\partial_{\tau}s^1 s^2 
- \partial_{\tau}s^2 s^1)+\frac{\rho_s}{2}\,\partial_{\mu}\OM\cdot
\partial_{\mu}\OM\right]. 
\end{equation} 
Here, $\rho_s=S^2 J/2$ is the spin stiffness.
This result was also obtained by Leutwyler in \cite{Leut94} using chiral 
perturbation theory. He showed that this theory has a non-relativistic
dispersion relation, attributed to the existence of a conserved order 
parameter.

Notice that the field $\OM(x)$ is a unit vector field. It therefore
lives in $S^2$, which is the coset
space corresponding to a symmetry breaking pattern
$SO(3)\to SO(2)$. Thus, the low-energy effective theory
in (\ref{eq:O(3)_eff_ac})
is a theory of Goldstone modes associated with 
this symmetry breaking. 

The mechanism of dimensional reduction was explained for 
the antiferromagnet in \cite{Has91}. Here, we adapt the discussion
for the ferromagnet. Consider a system of dimensions
$L\times L \times \beta$. For $L=\beta=
\infty$, the system is in the ordered ground state of the Heisenberg
ferromagnet. This breaks the
$SO(3)$ symmetry and, as discussed above, we obtain the 
low-energy theory of Goldstone bosons of 
(\ref{eq:O(3)_eff_ac}).
If we now consider the case in which the extent of the
additional dimension is taken to be finite, then
the Coleman-Hohenberg-Mermin-Wagner theorem tells us that there
cannot be massless excitations in a slab
\cite{MeWa66}. The Goldstone bosons must therefore 
pick up a small, non-perturbatively generated mass. 
One can use a block spin transformation to map
the 3-d $O(3)$-model in a slab of finite extent $\beta$ to a 2-d lattice 
$O(3)$ model. One averages the fields over blocks of size $\beta$ in the
third direction and size $a'$ in the two spatial
directions, decreasing the original cutoff $1/a$ down
to $1/a'$. To determine a suitable value for $a'$, consider the dispersion
relation of the ferromagnet, $E = (\rho_s/S) p^2$. 
This implies $1/\beta = (\rho_s/S) (1/{a'}^2)$,
and hence $a' = \sqrt{\rho_s \beta/S}$.
One thus obtains a 
two-dimensional lattice field theory, whose
degrees of freedom are the block-averaged fields 
living at the block centers. The lattice spacing of the 
new theory is therefore equal to $\sqrt{\rho_s \beta/S}$, 
which is different
from the lattice spacing of the original quantum Heisenberg 
model (see figure \ref{fig:blockspin}).
The correlation length of the 2-d $O(3)$ model in lattice
units is given by $\xi/a' \propto g^2 \exp\left( 2\pi/g^2\right)$.
The value of  $a'$ was found above, and we can identify the coupling
constant as $1/g^2 = \rho_s\beta$. We thus arrive at the expression
for the dependence of the correlation length on the extent
$\beta$ of the additional dimension,
\begin{equation} 
\xi\propto (\rho_s\beta)^{-1/2}\exp(2\pi\rho_s \beta). 
\end{equation} 
This relation was first found by Kopietz and Chakravarty in \cite{KoCh89},
where they used the same techniques of renormalization-group 
analysis that had been used for the antiferromagnet
in \cite{CHaN88} by Chakravarty, Halperin and Nelson.
The continuum limit is reached by taking the extent of
the additional dimension to infinity, since the
correlation length diverges exponentially as
$\beta\to\infty$. However, in this limit the extent
of the third dimension is much smaller than the
correlation length, i.e.\ $\xi \gg \sqrt{\rho_s \beta/S}$.
Thus, the fields are effectively constant in the 
$\tau$-direction, and the theory undergoes dimensional
reduction. 

\sfig{\mbox{\epsfig{figure=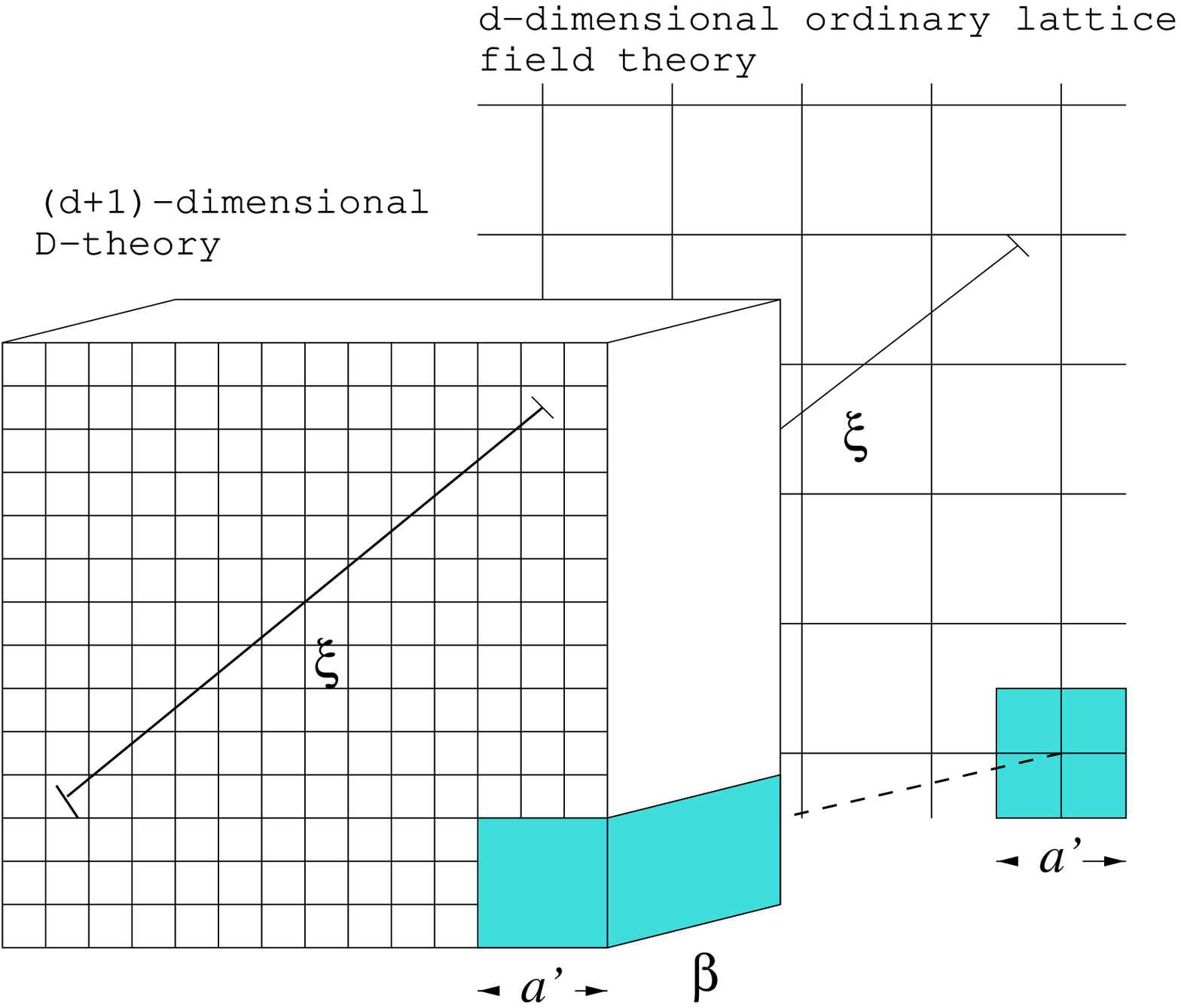,width=4in}}}{%
      TABLE ENTRY.   
      }{\sf%
      Dimensional Reduction of a D-theory: Averaging the
      $(d+1)$-dimensional effective field of the D-theory
      over blocks of size $\beta$ in the extra dimension
      and $a'$ in the physical directions results 
      in an effective $d$-dimensional Wilsonian lattice 
      field theory with lattice spacing $a'=\sqrt{\rho_s \beta/S}$.
      }{fig:blockspin}

\section{$U(N)$ Quantum Spins and Quantum Links}
\label{ch:U(N)}

In the following sections we will be considering models
whose fundamental degrees of freedom in the conventional
formulation are elements of unitary $N\times N$ matrices.
In D-theory these fields are replaced by quantum
operators, so that we have matrices whose
entries are operators rather than complex numbers.
However, we still want the Hamiltonian constructed from
these operators to be invariant under the same symmetries
as the conventional action. In particular, we will
be considering two models, the principal chiral model
with a global $SU(N)_L\times SU(N)_R\times U(1)_{L=R}$ symmetry, and Yang-Mills
theory with a local $SU(N)$ symmetry. In order to construct 
the appropriate D-theory Hamiltonians, let us consider
the operators associated with just one lattice site in the
case of the principal chiral model, or just one link in the
case of Yang-Mills theory. Dropping the site/link indices,
we denote these operators by $U^{ij}$. The appropriate
symmetries follow by construction if we also associate
with each site/link an $SU(N)_L\times SU(N)_R$ algebra,
generated by $\{ L^a\}$ and $\{ R^a\}$ which satisfy the
commutation relations
\begin{equation}
[L^a,L^b]=2if_{abc}L^c,\;\;\;
[R^a,R^b]=2if_{abc}R^c,\;\;\;
[L^a,R^b]=0.
\end{equation}
Here $f_{abc}$ are the usual structure constants of 
$SU(N)$. We then require that the site/link operator
variables transform as
\begin{equation}
\label{eq:trans_rl}
U'=\exp\left( -i \alpha^a L^a \right)
U\exp\left(i\beta^b R^b\right)
=\exp\left(i\alpha^a\lambda^a\right)U
 \exp\left(-i\beta^b\lambda^b\right),
\end{equation}
where the $\lambda^a$ are the Hermitian generators
of $SU(N)$ in the fundamental representation. These generators
satisfy
\begin{equation}
[\lambda^a,\lambda^b]=2if_{abc}\lambda^c,\;\;\;
\text{Tr}\lambda^a \lambda^b=2\delta^{ab}.
\end{equation}
The transformation rule (\ref{eq:trans_rl}) is implied by
the following commutation relations, 
\begin{equation}
[L^a,U^{ij}]=-\lambda^a_{ik} U^{kj},\;\;\;
[R^a,U^{ij}]=U^{ik}\lambda^a_{kj}.
\end{equation}
All of these commutation relations can be satisfied by
embedding the operators in an $SU(2N)$ algebra. In 
particular, the aforementioned $SU(N)_L\times SU(N)_R$ algebra
is embedded diagonally, while the $U^{ij}$ operators 
fill in the off-diagonal blocks.
To summarize, we get the full set of commutation relations
\begin{align}
&\left[ L^a,L^b\right] =2if_{abc}L^c,\;\;\; 
\left[ R^a,R^b\right] =2if_{abc}R^c, 
\notag \\
&\left[ R^a,U^{ij}\right] =U^{ik}\lambda^a_{kj},\;\;\; 
\left[ L^a,U^{ij}\right] =-\lambda^a_{ik}U^{kj},\;\;\;
\left[ T,U^{ij}\right] =2U^{ij}, 
\notag \\
&\left[ R^a,L^b\right] =
\left[ T,L^a\right]=\left[ T,R^a\right] =0, 
\notag \\
&\left[ \text{Re}\, U^{ij},\text{Re}\, U^{k\ell}\right]=
\left[ \text{Im}\, U^{ij},\text{Im}\, U^{k \ell}\right] 
=-i\left( \delta_{ik}\text{Im}\, \lambda^a_{j\ell}R^a
        +\delta_{j\ell}\text{Im}\,\lambda^a_{ik}L^a\right), 
\notag \\ 
&\left[ \text{Re}\, U^{ij},\text{Im}\, U^{k\ell}\right] =
i\left( \delta_{ik}\text{Re}\,\lambda^a_{j\ell} R^a
-\delta_{j\ell}\text{Re}\,\lambda^a_{ik}L^a+
\frac{2}{N}\delta_{ik}\delta_{j\ell} T \right).
\label{eq:pr_commrel}
\end{align}
\sfig{\mbox{\epsfig{figure=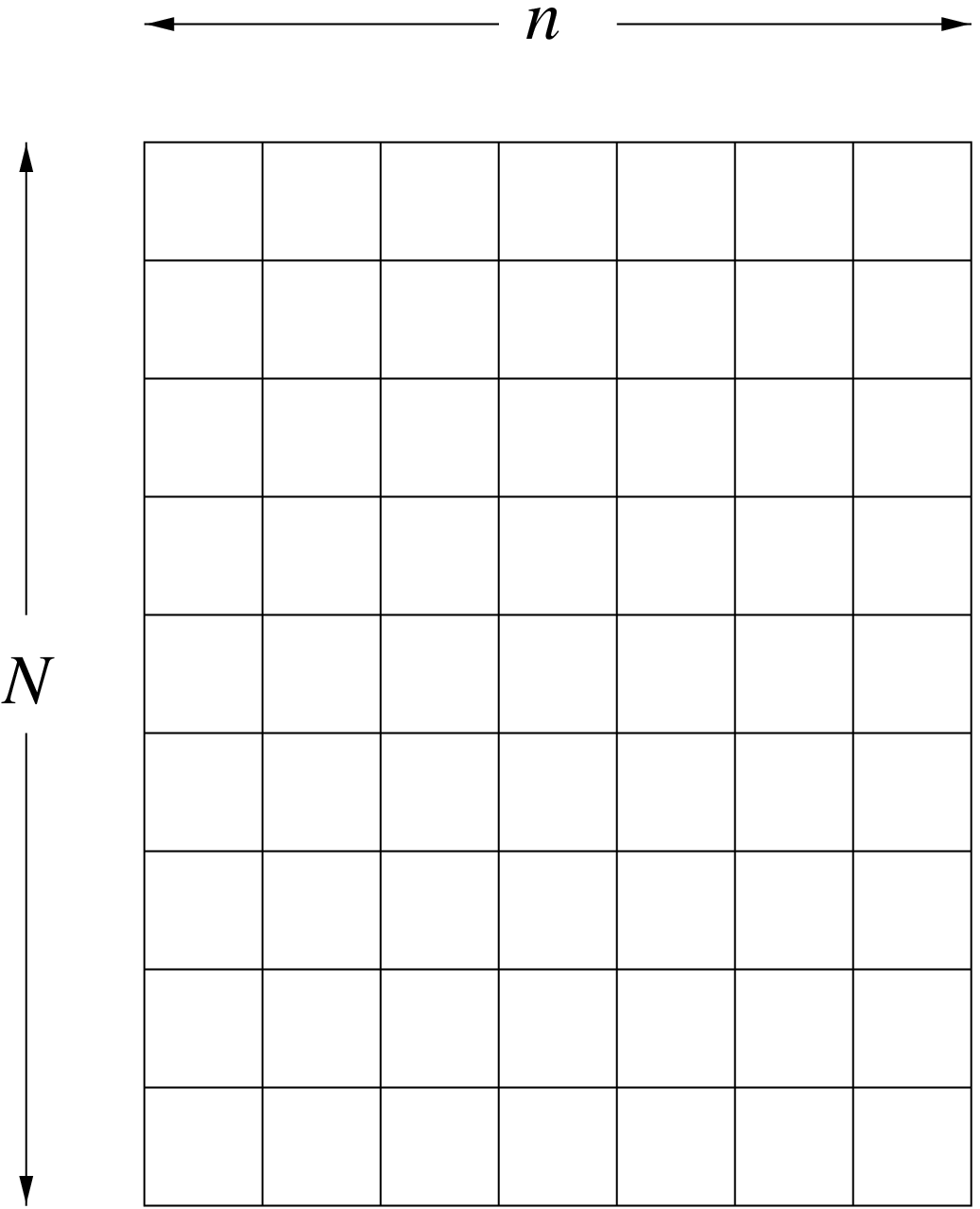,width=3in}}}{%
      TABLE ENTRY.   
      }{\sf%
      Young tableau of the $SU(2N)$ representation
      by which states at each lattice site/link transform.	
      }{fig:representations}
Here, $T$ generates an extra $U(1)$ symmetry. (Later, this
symmetry needs
to be broken explicitly to obtain an $SU(N)$ rather than $U(N)$
Yang-Mills theory.)
If we restrict ourselves to representations of $SU(2N)$
which correspond to rectangular Young tableaux
with $N$ rows and $n$ columns (as shown in figure \ref{fig:representations}), 
we can use 
a fermionic basis of rishons for our 
representation \cite{BChW97,ReSa89}, (we use the summation convention
for greek indices)
\begin{align}
&S^{ij}=L^{ij}=\left(\ell^{i\alpha\dagger}
\ell^{j\alpha}-\frac{n}{2}\delta^{ij}\right), &  
&S^{N+i,N+j}=R^{ij}=\left(r^{i\alpha\dagger}
 r^{j\alpha}-\frac{n}{2}\delta^{ij}\right), 
\notag \\
&S^{N+j,i}=-U^{ij}=
            -\ell^{i\alpha}r^{j\alpha\dagger}, &
&S^{j,N+i}=-\left(U^{\dagger}\right)^{ij}=
-\left(U^{ji}\right)^{\dagger}=-r^{i\alpha}
\ell^{j\alpha\dagger}, \notag \\
&T=\sum_i \left( r^{i\alpha\dagger}r^{i\alpha}
      -\ell^{i\alpha\dagger}\ell^{i\alpha}\right),& 
&\sum_{i}\left(\ell^{i\alpha\dagger}
\ell^{i\beta}\right. +
        \left. r^{i\alpha\dagger}r^{i\beta}\right)=
\delta^{\alpha \beta}N, \label{eq:2} 
\end{align} 
where $\alpha=1,\ldots, n$ is an additional rishon flavor index 
and $i,j=1,\ldots,N$. For convenience, we have chosen
these generators not to be traceless. We then have
$L^{a}=\lambda^{a}_{ij}L^{ij}$ and 
$R^{a}=\lambda^{a}_{ij}R^{ij}$.
The constraint (\ref{eq:2})
is needed to obtain the correct representation.

For the purpose of deriving systems of coherent states,
addressed in the next section, it
will be convenient to introduce the following notation,
\begin{equation}
c^{i\alpha}=\left\{\parbox{5cm}{$\begin{array}{ll}
\ell^{i\alpha}, & \mbox{for} \; 1\leq i\leq N; \\
r^{i\alpha}, & \mbox{for} \; N+1\leq i\leq 2N.
\end{array}$} \right.
\end{equation}
We then have 
\begin{equation}
S^{ij}=c^{i\alpha\dag} c^{j\alpha} -
\frac{n}{2}\delta^{ij}, \hspace{1cm}
\sum_i c^{i\alpha\dag} c^{i\beta}=\delta^{\alpha\beta}N.
\label{eq:pr01}
\end{equation}
Notice that the labels $i$ and $j$ now run from 1 to $2N$.

\section{Coherent States}
\label{ch:coherent}

In this section we first describe how to generate an over-complete
system of states, which we then use to set up a path integral.
Moreover, a Berry phase term is generated in the action and we work
out its form in terms of the degrees of freedom in the path integral.
These degrees of freedom are arranged into
$GL(N,\mathbb{C})$ matrices and we decompose such
a matrix into its Hermitian and unitary parts. This will allow
us to make contact with the respective target theories.

\subsection{Systems of Coherent States for $SU(2N)$}

Let us now construct a coherent state system for the types of
representations of $SU(2N)$  that we are interested in.
The general procedure is described in detail in
\cite{Pere86}. We must first pick a vector in the carrier space
that the chosen representation acts
in. It is convenient to choose a weight vector 
$\ket{\mu}$ as the initial
element of the coherent state system. 
We will see that for our choice of initial vector, the
corresponding isotropy subgroup, i.e.\ the subgroup of $SU(2N)$
transformations which leave the state defined by this vector
unchanged, is $SU(N)_L\times SU(N)_R\times U(1)_{L=R}$. 
This is a consequence of
the chosen type of representation with rectangular Young
tableaux of $N$ rows. Coherent states are then characterized by the elements 
of $X=SU(2N)/[SU(N)_L\times SU(N)_R\times U(1)_{L=R}]$.

We now proceed to constructing a concrete set of coherent 
states for representations of $SU(2N)$ with rectangular
Young tableaux of the kind previously described.
This discussion follows the steps outlined in \cite{ReSa89}.
Our basis of generators is given by (\ref{eq:pr01}). 
We can choose the Cartan
sub-algebra to be spanned by the following set of generators,
\begin{equation}
S^{ii}=\sum_{\alpha}c^{i\alpha\dag}c^{i\alpha}-\frac{n}{2},
\end{equation}
for $i=1,\ldots,2N$. Notice that (\ref{eq:pr01}) imposes
one constraint on this set of generators, consistent with 
the fact that $SU(2N)$ has rank $2N-1$. The remaining operators
$S^{ij}$ with $i\neq j$ are the ``raising'' and
``lowering'' operators which complete the canonical
Cartan basis for the Lie algebra.

For the given representation we obtain the highest-weight
vector as follows,
\begin{equation}
\label{eq:co04}
\ket{\psi_0}={\cal C}\left[\varepsilon^{ab\cdots}
c^{a\alpha\dag}c^{b\alpha\dag}\cdots\right]
\left[\varepsilon^{cd\cdots}c^{c\beta\dag}c^{d\beta\dag}\cdots
\right]
\cdots \ket{0},
\end{equation} 
where there are $N$ creation operators in each square 
bracket, and there are $n$ square bracketed terms all 
together. The indices $a,b,\ldots$ run through all values 
1 to $N$, while $\alpha,\beta,\ldots$ run from 1 to $n$. 
We are symmetrizing the row indices of the 
Young tableau, and antisymmetrizing the column indices. The 
normalization constant 
${\cal C}$ is chosen so that $\bracket{\psi_0}{\psi_0}=1$.
The weight of this state is given by
\begin{equation}
S^{ii} \ket{\psi_0}=\left\{\parbox{6.5cm}{$\begin{array}{ll} 
\frac{n}{2}\ket{\psi_0}, &\mbox{for} \; 1\leq i\leq N; \\
-\frac{n}{2}\ket{\psi_0}, &\mbox{for} \; N+1\leq i\leq 2N.
\end{array}$}\right.
\end{equation}
We can obtain a coherent state system by applying all 
possible group transformations modulo the isotropy subgroup 
of $\ket{\psi_0}$ to 
our chosen initial vector. In terms of the chosen basis
for the Lie algebra we obtain a group element of $SU(2N)$
by exponentiating an anti-Hermitian combination of generators,
and so we have,
\begin{equation}
\label{eq:co01}
\ket{q}=\exp\left(-q^{ij} S^{ji}+q^{ij*} S^{ij} \right)
\ket{\psi_0},
\end{equation}   
where the index $j$ runs through the values 1 to $N$, and 
$i$ runs from $N+1$ to $2N$. It is easily checked
that if $i$ and $j$ were to fall in the same range of values,
i.e.\ either both are between 1 and $N$ or both are between
$N+1$ and $2N$, then $\ket{q}$ would just be equal (or in the 
case $i=j$ proportional) to $\ket{\psi_0}$. 
Hence, the isotropy subgroup of the vector $\ket{\psi_0}$ is
$SU(N)_L\times SU(N)_R\times U(1)_{L=R}$ as mentioned above. The $q^{ij}$ are
$N^2$ independent complex numbers, which can be arranged
into a $GL(N,\mathbb{C})$ matrix. Under $SU(2N)$ rotations, the
generators $S^{ij}$ transform in the adjoint representation,
\begin{multline}
\exp\left( q^{k\ell}S^{\ell k}-q^{k\ell *}S^{k\ell}\right) S^{ij}
\exp\left( -q^{mn}S^{nm}+q^{mn*}S^{mn} \right) \\
= \left[ \exp \mat{0}{q^{\dag}}{-q}{0}\right]^{ik} S^{k\ell}
\left[\exp \mat{0}{-q^{\dag}}{q}{0}\right]^{\ell j}.
\label{eq:adjoint}
\end{multline}
The states $\ket{q}$ 
are normalized to one, and from (\ref{eq:adjoint}) they clearly
satisfy the following important identity
\begin{equation}
\bra{q}S^{ij}\ket{q}=\frac{n}{2}Q^{ij},
\end{equation}
where $Q$ is given by
\begin{equation} 
\label{eq:1}
Q=\exp\left[\mat{0}{q^{\dag}}{-q}{0}\right] 
\mat{\1_N}{0}{0}{-\1_N}
\exp\left[\mat{0}{-q^{\dag}}{q}{0}\right].
\end{equation}
In particular, if we write the matrix $Q$ as 
\begin{equation}
Q=\mat{\ell}{w^{\dag}}{w}{r},
\end{equation}
we see  that 
\begin{equation}
\label{qmatrixel}
\bra{q}U^{ij}\ket{q}=-\frac{n}{2}w^{ij}.
\end{equation}

From the discussion of how to set up a coherent state path
integral in the case of the Heisenberg model, we know that
a Berry phase term of the form
\begin{equation}
S_B=\int_0^{\beta}d\tau\left(
\frac{\bracket{q(\tau+\varepsilon)}{q(\tau)}-1}{\varepsilon}
\right).
\end{equation}
is generated at each site/link as part of the action. 
To manipulate this term,
notice that for any operator $M$,
we have \cite{Feyn51},
\begin{equation}
\frac{d}{dx}e^M =\int_0^1 dv\,e^{M(1-v)}\frac{dM}{dx}e^{Mv}.
\end{equation} 
Using the expression for $\ket{q}$ found in (\ref{eq:co01}),
we can write the Berry phase term as,
\begin{eqnarray}
S_B&=&-\int_0^{\beta}d\tau \bra{q(\tau)}\frac{d}{d\tau}
    \ket{q(\tau)}
 \notag \\
 &=&-\int_0^{\beta} \bra{\psi_0}\exp(q^{ij}S^{ji}-
q^{ij*}S^{ij})\frac{d}{d\tau}\exp(-q^{ij} S^{ji}+
q^{ij*}S^{ij})\ket{\psi_0} \notag \\
 &=&-\int_0^{\beta}d\tau \int_0^1 dv \,\bra{\psi_0}
\exp[-v(-q^{ij}S^{ji}+q^{ij*}S^{ij})] \notag \\
 & & \hspace{1.3cm}
  \times\left(-\frac{\partial q^{ij}}{\partial \tau}S^{ji}+
\frac{\partial q^{ij*}}{\partial \tau}S^{ij}\right)
\exp[v(-q^{ij} S^{ji}+q^{ij*}S^{ij})]\ket{\psi_0}. 
\end{eqnarray}
If we now define 
\begin{equation}
\bra{vq(\tau)}S^{ij}\ket{vq(\tau)}\equiv 
\frac{n}{2}Q^{ij}(\tau,v),
\end{equation}
we can simplify the above to
\begin{eqnarray}
S_B&=&-\frac{n}{2}\int_0^{\beta}d\tau\int_0^1 dv
\left( -\frac{\partial q^{ij}}{\partial \tau}Q^{ji}(\tau,v)
+\frac{\partial q^{ij*}}{\partial \tau}Q^{ij}(\tau,v) 
\right) \notag \\
 &=&-\frac{n}{2}\int_0^{\beta}d\tau\int_0^1 dv\,
\mbox{Tr}\left[\mat{0}{\frac{\partial q^{\dag}}{\partial\tau}}
{-\frac{\partial q}{\partial \tau}}{0} Q(\tau,v)
\right].
\label{eq:co02}
\end{eqnarray}
As a function of $v$ the matrix $Q(\tau,v)$ now satisfies
\begin{equation}
Q(\tau,0)=\mat{\1_N}{0}{0}{-\1_N},
\end{equation}
and $Q(\tau,1)\equiv Q(\tau)$. Integrating (\ref{eq:co02})
by parts gives
\begin{equation}
S_B=\frac{n}{2}\int_0^{\beta}d\tau \int_0^1 dv\,
\mbox{Tr}\left[\mat{0}{q^{\dag}(\tau)}{-q(\tau)}{0}
\frac{\partial}{\partial\tau}Q(\tau,v)\right].
\end{equation}

It is not hard to see that
\begin{equation}
\mat{0}{q^{\dag}(\tau)}{-q(\tau)}{0}=-\frac{1}{2}
Q(\tau,v)\frac{\partial Q(\tau,v)}{\partial v},
\end{equation}
and this leads to the final result for the Berry phase,
\begin{equation}
S_{B}=-\frac{n}{4}\int_{0}^{\beta} d\tau \int_{0}^{1}
d v 
\left[\mbox{Tr}\left( Q(\tau,v)
\frac{\partial Q(\tau,v)}{\partial v}
\frac{\partial Q(\tau,v)}{\partial \tau}\right)\right] 
\end{equation}

In the above derivation we used a 
specific dependence of $Q(\tau,v)$ on the variable $v$,
which satisfies the boundary conditions
\begin{gather}
Q(\tau,0)=Q(\tau',0), \quad \text{for all}
                               \;\tau,\tau'; \notag \\
Q(\tau,1)=Q(\tau);  \quad
Q(0,v)=Q(\beta ,v).\label{eq:co08}
\end{gather}
Thus, the field $Q(\tau,v)$ lives in a rectangle 
$0\leq \tau\leq \beta$ and $0\leq v\leq 1$.
From the periodic boundary conditions in the $\tau$ direction,
we can interpret $Q(\tau)$ as defining a closed curve, 
parametrized by $\tau$, and with the parameter $v$ filling in
the space enclosed by the curve to form a disc in the Grassmann manifold
$G(N,2N)=SU(2N)/[SU(N)_L\times SU(N)_R\times U(1)_{L=R}]$. 
In \cite{ReSa89} it was shown
that $S_B$ is independent of the particular surface that has 
this boundary, up to multiples of $2\pi n ki$ for 
$k\in \mathbb{Z}$. This result was derived as a direct consequence
of the fact that the second homotopy group of this Grassmann 
manifold is just $\Pi_2(G(N,2N))=\mathbb{Z}$, the group of integers.

Consider now the $GL(N,\mathbb{C})$ matrix $q$
of (\ref{eq:co01}). As shown in the appendix, we can decompose such a 
matrix into the product of a left-coset Hermitian matrix $b$ and
a unitary matrix $u$, $q=bu$. 
Upon substituting this decomposition into
(\ref{eq:1}), we obtain
\begin{equation}
\label{eq:cosetdecomp}
Q=\mat{u^{\dag}\cos(2b)u}{-u^{\dag}\sin(2b)}{-\sin(2b)u}
{-\cos(2b)}.
\end{equation} 

In order to use this result to simplify the Berry phase term in the 
action, we represent $S_B$ as
\begin{equation}
\label{eq:pr_l_berry}
S_B=\frac{n}{8}\int d^2 \xi\, \varepsilon_{pq}
\mbox{Tr}\,[Q\partial_p Q\partial_q Q],
\end{equation}
where $p,q$ take the values 1, 2, and $\xi_1=\tau$,
and $\xi_2=v$, and the integral is over a rectangle in
$(\tau,v)$ space. We parametrize $Q(t,v)$ in the following way, 
\begin{align}
Q&(\tau,v)= \notag \\
&\exp\left[\mat{0}{q^{\dag}(\tau,v)}{-q(\tau,v)}{0}
\right]
\mat{\1_N}{0}{0}{-\1_N}
\exp\left[\mat{0}{-q^{\dag}(\tau,v)}{q(\tau,v)}{0}
\right],
\end{align}
where $q(\tau,v)$ is a smooth function on the rectangle,
such that the boundary conditions (\ref{eq:co08})
are satisfied. We can then decompose the matrix 
$q(\tau,v)$ as before,
\begin{equation}
q(\tau,v)=b(\tau,v)u(\tau,v),
\end{equation}
and find 
\begin{equation}
Q=\mat{u^{\dag}(\tau,v)\cos(2b(\tau,v))u(\tau,v)}
{-u^{\dag}(\tau,v)\sin(2b(\tau,v))}
{-\sin(2b(\tau,v))u(\tau,v)}
{-\cos(2b(\tau,v))}.
\end{equation}
Substituting this expression into the integrand of 
(\ref{eq:pr_l_berry}), we find that after some algebra
it reduces to
\begin{equation}
\varepsilon_{pq}\mbox{Tr}\left[Q\partial_p Q\partial_q Q
\right]
=-4\varepsilon_{pq}\mbox{Tr}\left[ \partial_q\left( 
\cos(2b(\tau,u))u(\tau,u)\partial_p u^{\dag}(\tau,u)
\right)\right].
\end{equation}
Hence, the Berry phase term simplifies to
\begin{eqnarray}
S_B&=&- \frac{n}{2}\int_0^{\beta}d\tau \int_0^1 dv
\left(\partial_v \left\{ \mbox{Tr}\left[
\cos(2b)u\partial_{\tau}
u^{\dag}\right]\right\} -\partial_{\tau}\left\{ 
\mbox{Tr}\left[\cos(2b)
u\partial_v u^{\dag}\right]\right\}\right) \notag \\
 &=&-\frac{n}{2}\int_0^{\beta}d\tau \,\mbox{Tr}
\left[\cos(2b)u\partial_{\tau}u^{\dag}\right],
\label{eq:berry1}
\end{eqnarray}
where we have used the boundary conditions on 
$b(\tau,v)$ and $u(\tau,v)$ to obtain the last line.

\section{Principal Chiral Model}
\label{ch:principal}

In this section we use symmetry considerations to formulate the principal 
chiral model as a quantum spin model. We then set up a coherent state
path integral using the results of the previous section. The resulting
Lagrangian is expanded around its minimum to obtain a 3-d principal 
chiral model as the low-energy effective theory of the quantum spin model.
Finally, we explain how the mechanism of dimensional reduction gives
rise to the 2-d target theory.

\subsection{D-Theory Formulation}

The action of the 2-d principal chiral model in the continuum
is given by
\begin{equation} 
S[u]=\frac{1}{2g^2}\int d^2 x\, \mbox{Tr}
       \left[\partial_{\mu}u^{\dagger}(x)
       \partial_{\mu}u(x)\right], 
\end{equation}
where  the $u(x)$ are unitary $N\times N$ matrices.
In Wilson's approach to regularizing the theory, space-time is
discretized by introducing a regular lattice. 
Derivatives are replaced by finite 
differences to obtain an action of the form
\begin{equation}
S[u]=
-\frac{1}{g^2}\sum_{<xy>} \mbox{Tr}\left[u_{x}^{\dagger}
u_{y}\right].
\end{equation}
The target theory has a
global $SU(N)_{L}\times SU(N)_{R}\times U(1)_{L=R}$ symmetry of the
form $u_x\to u_x'=Lu_x R^{\dagger}$, where 
$L$ and $R$ are unitary matrices. It is known that
this symmetry breaks to an $SU(N)$ vector symmetry ($L=R$)
at $g=0$. Due to the Mermin-Wagner theorem, however,
the symmetry cannot break for $g>0$.

Let us now replace the classical fields $u^{ij}_x$
by quantum operators $U^{ij}_x$ and write down a 
D-theory Hamiltonian, which evolves the 
two-dimensional system in an additional Euclidean 
time direction,
\begin{equation}
H=2J\sum_{x,\mu}\text{Re}\,\text{Tr}\,U_x U^{\dag}_{x+\hat{\mu}}
=J\sum_{x,\mu}\left[U_{x}^{ij}
\left(U_{x+\hat{\mu}}^{ij}\right)^{\dagger}+U_{x+\hat{\mu}}^{ij}
\left(U_{x}^{ij}\right)^{\dagger}
\right].
\end{equation}
We would like this Hamiltonian to have an
$SU(N)_{L}\times SU(N)_{R}\times U(1)_{L=R}$ symmetry, i.e.\ 
$\left[ G^a_L,H\right]=\left[ G^a_R,H \right] =[T,H]=0$,
where $G_L^a$ and  $G_R^a$ are  mutually commuting 
sets of $SU(N)$ generators and $T$ generates a $U(1)$ symmetry.
As we saw in chapter \ref{ch:U(N)}, this can be realized
by embedding $SU(N)_L\times SU(N)_R\times U(1)_{L=R}$ diagonally in
$SU(2N)$. In particular, we have an $SU(2N)$ algebra of the
form (\ref{eq:2}) at each lattice site.
Labeling the generators that correspond to the lattice site $x$
by a subscript $x$, we can write the generators of the
algebra of the global $SU(N)_L\times SU(N)_R\times U(1)_{L=R}$ symmetry
group as $G_L^a=\sum_x L_x^a$, $G_R^a=\sum_x R_x^a$ and
$T=\sum_x T_x$.
It then follows that $[G_L^a,H]=[G_R^a,H]=[G_L^a,G_R^b]=[T,H]=0$.

For $J<0$ this model is ferromagnetic.
We choose the same representation of $SU(2N)$ for the generators
at each site of the lattice, namely the one mentioned above, with
a rectangular Young tableau as shown in figure
\ref{fig:representations}.
Note that the properties of the system defined by the Hamiltonian $\,H$ are 
completely determined, once the 
representation of 
$SU(2N)$ has been specified.

\subsection{The Continuum Limit of the Theory}
\label{sec:contlim}

Using the results of chapter \ref{ch:coherent}, we set up
a coherent state path integral for the partition function.
The Berry phase term of the corresponding action was
calculated in (\ref{eq:berry1}). 
The other term in the action is given by
\begin{eqnarray}
\lefteqn{\int_0^{\beta}d\tau \,{\cal H}(Q(\tau))}  \notag \\
 &=&\frac{Jn^2}{4}\int_0^{\beta}d\tau \sum_{x,\mu}
\left[ w^{ij}_x
\left(w^{ij}_{x+\hat{\mu}}\right)^{*}
+\left(w^{ij}_{x+\hat{\mu}}\right)^{*}
w^{ij}_{x} \right] \notag \\ 
 &=&\frac{Jn^2}{4}\int_0^{\beta}\sum_{x,\mu}\text{Tr}
\left[w_{x}w^{\dag}_{x+\hat{\mu}}+w_{x+\hat{\mu}}w^{\dag}_{x}
\right]. 
\label{eq:prhamilton}
\end{eqnarray}
As discussed above, we have $w=-\sin(2b)u$ and we 
define $s\equiv \sin(2b)$. Then $w=-su$, where
$s=s^{\dag}$. We can think of $s$ as the radial component
and of $u$ as the phase of the matrix $w$. Due to the
sine function and the fact that we can cover the
coset space $SU(2N)/[SU(N)_L\times SU(N)_R\times U(1)_{L=R}]$ 
by limiting the matrix
$b$ to have eigenvalues between $0$ and $\pi/2$, the 
eigenvalues of $s$ are constrained to lie between 0 and 1.
Substituting the above coset decomposition into 
(\ref{eq:prhamilton}), we obtain
\begin{align}
\frac{Jn^2}{4}\int_0^{\beta}&d\tau\sum_{x,\mu}\text{Tr}
\left[ s_{x}u_{x}u^{\dag}_{x+\hat{\mu}}s_{x+\hat{\mu}}+ 
       s_{x+\hat{\mu}}u_{x+\hat{\mu}}u^{\dag}_{x}s_{x}\right] \notag \\
&=-\frac{Jn^2}{4N}\int_0^{\beta}d\tau \sum_{x,\mu}\text{Tr} 
   \bigg[\frac{1}{2}
   (s_{x}s_{x+\hat{\mu}}-
   s_{x+\hat{\mu}}s_{x})
   (u_{x}u^{\dag}_{x+\hat{\mu}}
    -u_{x+\hat{\mu}}u^{\dag}_{x})
   \notag \\
& \hspace{1cm}
   +\frac{1}{4}(2+u_{x}u^{\dag}_{x+\hat{\mu}}+u_{x+\hat{\mu}}u^{\dag}_{x}) 
   (s_{x+\hat{\mu}}-s_{x})
   (s_{x+\hat{\mu}}-s_{x}) \notag \\
& \hspace{1cm}
   +\frac{1}{4}(s_{x+\hat{\mu}}+s_{x})(s_{x+\hat{\mu}}+s_{x}) 
   (u_{x+\hat{\mu}}-u_{x})
   (u_{x+\hat{\mu}}^{\dag}-u^{\dag}_{x})
   -2s_{x}s_{x}\bigg]. 
\label{eq:pr_expandterms}
\end{align}

We would like to expand this action around its minimum.
Since the eigenvalues of $s$ are bounded by 1, the minimum
of the action occurs when $s=\1$ (up to a global $SU(N)_L\times
SU(N)_R\times U(1)_{L=R}$ rotation), and $u$ is constant across the
lattice. We now introduce fluctuations in the fields, defining
the forward lattice derivative,
\begin{equation}
\Delta^f_{\mu}u_x = \frac{u_{x+\hat{\mu}}-u_{x}}{a},
\end{equation}
and writing
\begin{equation}
s_{x}=\sin(2b_{x}) 
    =\sin\left(2\left(\frac{\pi}{4}+a\epsilon_{x}\right)\right)
    \approx \1 -2a^2 \epsilon^2_{x}.
\end{equation}
Substituting these expressions into (\ref{eq:pr_expandterms}), 
expanding to quadratic order in $a$  and dropping an irrelevant
constant, we obtain
\begin{equation}
S=S_B -\frac{Jn^2}{4}\int_0^{\beta}d\tau \sum_{x,\mu}a^2 \,\text{Tr}
\left[\left(\Delta^f_{\mu}u_{x}\right)\left(\Delta^f_{\mu}u^{\dag}_{x}\right)
+8 \epsilon^2_{x}
\right].
\end{equation}
Next, we turn to the Berry phase term, and expand the
fields in the same way as before,
\begin{eqnarray}
S_B&=&-\sum_{x} \frac{n}{2}\int_0^{\beta}
d\tau\,
\mbox{Tr}\left(\cos(2b_{x})u_{x}\partial_{\tau}u_{x}^{\dag}\right)
\notag \\
&\approx&-\sum_x \frac{n}{2}\int_0^{\beta}d\tau\, \text{Tr}
\left[\left( \cos\left(\frac{\pi}{2}\right)
-2a\epsilon_{x}\sin\left(\frac{\pi}{2}\right) \right)
u_{x}\partial_{\tau}u^{\dag}_{x}\right] \notag \\
&=&\sum_x \frac{n}{2}\int_0^{\beta}d\tau\, \text{Tr}\left[
2a\epsilon_{x}u_{x}\partial_{\tau}u^{\dag}_{x}\right]
\end{eqnarray}
We thus obtain the final expression for the
action in the continuum limit,
\begin{eqnarray}
S&=&
\int_0^{\beta}d\tau\int d^2 x\, \mbox{Tr} \left[  
-\frac{Jn^2}{4}\partial_{\mu}
u(x) \partial_{\mu} u^{\dag}(x)
-4Jn^2\epsilon^2(x)
+\frac{n}{a}\epsilon(x)u(x)\partial_{\tau}u^{\dag}(x)
\right] \notag \\
&=&\int_0^{\beta}d\tau \int d^2 x \,\frac{\rho_s}{2}\text{Tr}\left[
\partial_{\mu}u\partial_{\mu}u^{\dag}+\frac{1}{c^2}\partial_{\tau}u
\partial_{\tau}u^{\dag}
\right].
\end{eqnarray}
In this equation, we have integrated out the shifted field
\begin{equation}
\epsilon'(x)=\epsilon(x)-\frac{1}{8Jna}u(x)\partial_{\tau}u^{\dag}(x).
\end{equation}
The spin stiffness is given by $\rho_s=|J|n^2/2$,
and $c=2 n a |J|$ is the spin wave
velocity. 

Notice that we now have a theory with a relativistic dispersion relation.
The ferromagnetic $SO(3)$ spin model has a non-relativistic dispersion
relation \cite{Leut94}, because the order parameter commutes with the
Hamiltonian. In the ferromagnetic principal chiral 
model on the other hand, the order parameter 
$U^{ij}=\sum_x U_x^{ij}$ does not commute with $H$, so a 
relativistic dispersion relation comes as no surprise.

The  three-dimensional system will dimensionally reduce
if the correlation length is much larger than the
extent of the third dimension, $\xi\gg \beta c$. 
If we now assume that 
this is the case, then the fields $u$ will have
no dependence on $\tau$, and the integration over
$\tau$ becomes trivial,
\begin{equation}
S=\frac{\beta\rho_s}{2}\int d^2 x \, \text{Tr}
\left[ \partial_{\mu}u \partial_{\mu}u^{\dag}
\right].
\end{equation}
From \cite{Wieg84} we know that the correlation for
the two-dimensional principal chiral model is given by
\begin{equation}
\label{eq:KoCh}
\xi\propto \exp\left( \frac{2\pi}{g^2 N}\right)
=\exp\left( \frac{2\pi \beta\rho_s}{N}\right).
\end{equation}
When performing a blockspin transformation in the way described for
the Heisenberg ferromagnet, the new lattice spacing for a system with
a relativistic dispersion relation, such as the present one, is
$a' = \beta c$. Equation (\ref{eq:KoCh})
is consistent with $\xi\gg \beta c$ in the
zero temperature $(\beta\to\infty)$ limit, so dimensional
reduction does indeed occur.

\section{Gauge Theory}
\label{ch:gauge}

The quantum link formulation of Yang-Mills theory
was worked out in \cite{BChW97} and we shall
review it here, before showing how it is related to
the classical formulation of the theory. 
First, recall Wilson's action for lattice gauge
theory with gauge group $SU(N)$ \cite{Wils74},
\begin{equation}
S[u]=-\frac{1}{g^2}\sum_{x,\mu\neq\nu} 2\text{Re }\text{Tr}
\left[ u_{x,\mu}u_{x+\hat{\mu},\nu} 
u_{x+\hat{\nu},\mu}^{\dag}
u_{x,\nu}^{\dag}\right].
\end{equation}
Here, $x$ labels the sites of a 4-d hyper-cubic
lattice, and
the $u_{x,\mu}$ are $SU(N)$ matrices, associated
with each link $(x,\mu)$ on the lattice.
This action is invariant under local $SU(N)$
transformations of the form
$u_{x,\mu}\to u_{x,\mu}'$, where
\begin{equation}
u_{x,\mu}'=\exp\left( i \alpha_x^a 
\lambda^a \right) u_{x,\mu}
\exp\left( -i \alpha_{x+\hat{\mu}}^b
\lambda^b \right).
\end{equation}
The classical partition function for this system
is given by
\begin{equation}
Z=\int {\cal D}u\, \exp\left( -\frac{1}{g^2}S[u]
\right).
\end{equation}

In the D-theory formulation 
we replace the classical fields that make up
the entries $u_{x,\mu}^{ij}$ of the $u_{x,\mu}$ matrices in the action 
by quantum operators $U_{x,\mu}^{ij}$, to obtain
a quantum Hamilton operator that evolves the
system in an additional Euclidean direction.
The Hamilton operator takes the form
\begin{equation}
H=J\sum_{x,\mu\neq\nu}
\left[ U_{x,\mu}^{ij}U_{x+\hat{\mu},\nu}^{jk}
\left(U_{x+\hat{\nu},\mu}^{\ell k}\right)^{\dag}
\left(U_{x,\nu}^{i\ell}\right)^{\dag} +\text{h.c.}\right]
+J'\sum_{x,\mu}[\det U_{x,\mu}+\det U_{x,\mu}^{\dag}]. 
\end{equation}
The determinant term is understood to mean
\begin{equation}
\det U_{x,\mu}=\frac{1}{N!}
\epsilon_{i_1 i_2\cdots i_N} \epsilon_{j_1 j_2\cdots
j_N} U_{x,\mu}^{i_1 j_1}U_{x,\mu}^{i_2 j_2}
\cdots U_{x,\mu}^{i_N j_N}.
\end{equation}
It has been introduced into the Hamiltonian
to break an extra $U(1)$ symmetry that would
otherwise be present and lead to a $U(N)$ rather
than an $SU(N)$ gauge invariant model.

This Hamilton operator has to be invariant under
gauge transformations, i.e., we require that
\begin{equation}
[H, G^a_x]=0,
\end{equation}
where $G^a_x$ are the generators of an
$SU(N)$ algebra at each lattice site $x$, 
obeying the commutation relations
\begin{equation}
[G_x^a, G_y^b]=2i \delta_{xy}f_{abc}
G_x^c.
\end{equation}
In the Hilbert space, a general gauge transformation 
is represented by the operator
$\prod_x \exp(i\alpha_x^aG_x^a)$.
We can construct gauge covariant transformations
of the fields by requiring that
\begin{align}
U_{x,\mu}'&= \prod_{y} \exp\left( -i \alpha_y^a
 G_y^a\right) U_{x,\mu}\prod_{z}
\exp\left( i \alpha_z^b G_z^b\right) 
\notag \\ 
 &= \exp\left( i \alpha_x^a  \lambda^a
\right) U_{x,\mu} \exp\left(
-i\alpha_{x+\hat{\mu}}^b
 \lambda^b\right).
\end{align}
This implies commutation relations of the form
\begin{equation}
[G_x^a, U_{y,\mu}]=\delta_{x,y+\hat{\mu}}
U_{y,\mu}\lambda^a 
-\delta_{x,y}\lambda^a U_{y,\mu}.
\end{equation}
In order to satisfy these relations, we write
\begin{equation}
G_x^a=\sum_{\mu}(R^a_{x-\hat{\mu},\mu}+
L^a_{x,\mu}),
\end{equation}
where $R^a_{x,\mu}$ and $L^a_{x,\mu}$ are 
generators of left and right gauge transformations
of the link variable $U_{x,\mu}$. They generate
an $SU(N)_R\times SU(N)_L$ algebra on each link,
which can be embedded diagonally in
the algebra of $SU(2N)$, with the commutation 
relations as given in (\ref{eq:pr_commrel}).

We choose representations for the $SU(2N)$ algebra
with rectangular Young tab\-le\-aux, as we already
did in the case of the principal chiral model.
In particular, we can use the rishon representation
of (\ref{eq:2})-(\ref{eq:pr01}).
In contrast to the principal chiral model, 
operators now live on
the links and not on the lattice sites. 
The notation is the following,
\begin{align}
&S^{ij}_{x,\mu}=L^{ij}_{x,\mu}=
       \left(c^{i\alpha\dagger}_{x,\mu}
c^{j\alpha}_{x,\mu}-\frac{n}{2}\delta^{ij}\right), &
&S^{N+i,N+j}_{x,\mu}=R^{ij}_{x,\mu}=
\left( c^{i\alpha\dagger}_{x+\hat{\mu},-\mu}
c^{j\alpha}_{x+\hat{\mu},-\mu}
-\frac{n}{2}\delta^{ij}\right), 
\notag \\
&S_{x,\mu}^{N+j,i}=-U^{ij}_{x,\mu}=
         -c^{i\alpha}_{x,\mu}
     c^{j\alpha\dagger}_{x+\hat{\mu},-\mu}, &
&S_{x,\mu}^{j,N+i}
=-\left(U_{x,\mu}^{ji}\right)^{\dagger}=
-c^{i\alpha}_{x+\hat{\mu},-\mu}
c^{j\alpha\dagger}_{x,\mu}, \notag \\
&T_{x,\mu}=\sum_i
( c^{i\alpha\dagger}_{x+\hat{\mu},-\mu}
c^{i\alpha}_{x+\hat{\mu},-\mu}
-c^{i\alpha\dagger}_{x,\mu}
c^{i\alpha}_{x,\mu}), 
& &\sum_{i}\bigl(c_{x,\mu}^{i\alpha\dagger}
c_{x,\mu}^{i\beta} +
    c_{x+\hat{\mu},-\mu}^{i\alpha\dagger}
c_{x+\hat{\mu},-\mu}^{i\beta}\bigr)=
\delta^{\alpha \beta}N, \label{eq:ym_rishon2} 
\end{align} 
where $\alpha=1,\ldots,n$ and $i,j=1,\ldots,N$.

The next step is to set up a coherent state path
integral as discussed in chapter
\ref{ch:coherent}. 
We will consider the analog of (\ref{qmatrixel})
and the coset decomposition of 
(\ref{eq:cosetdecomp}) to determine some
properties of the matrix $w$ that appears in
the action of the coherent state path integral.
We have
\begin{equation}
U^{ij}_{x,\mu}=-c_{x+\hat{\mu},-\mu}^{ja\dag}
c_{x,\mu}^{ia}.
\end{equation}
From (\ref{qmatrixel}) we obtain,
\begin{equation}
\langle q| U^{ij}_{x,\mu}| q\rangle =
-\frac{n}{2}w^{ij}_{x,\mu}.
\end{equation}
Now consider
\begin{align}
-\frac{n}{2}w^{ij}_{x+\hat{\mu},-\mu} &=
\langle q| U^{ij}_{x+\hat{\mu},-\mu}| q\rangle 
= -\langle q| c_{x,\mu}^{ja\dag}
c_{x+\hat{\mu},-\mu}^{ia}|q\rangle \notag \\
&= \langle q| \left( U_{x,\mu}^{ji}\right)^{\dag}
|q\rangle 
= \langle q| U_{x,\mu}^{ji}
|q\rangle^{*} 
= -\frac{n}{2}w^{ji*}_{x,\mu},
\end{align}
and hence we see that
\begin{equation}
w_{x+\hat{\mu},-\mu}=w^{\dag}_{x,\mu}.
\label{eq:qtransf}
\end{equation}
The coset decomposition is $w_{x,\mu}
=-s_{x,\mu}u_{x,\mu}$, where $s=s^{\dag}$
and $uu^{\dag}=\1$. Taken together with
(\ref{eq:qtransf}), this leads to 
\begin{align}
w_{x+\hat{\mu},-\mu}&= -s_{x+\hat{\mu},-\mu}
u_{x+\hat{\mu},-\mu} = w^{\dag}_{x,\mu} \notag \\
&= -u^{\dag}_{x,\mu}s_{x,\mu} 
= -u^{\dag}_{x,\mu}s_{x,\mu}u_{x,\mu}u^{\dag}_{x,\mu},
\end{align}
and we deduce that
\begin{equation}
u_{x+\hat{\mu},-\mu}= u^{\dag}_{x,\mu}, \hspace{0.5cm}
s_{x+\hat{\mu},-\mu}= u^{\dag}_{x,\mu} \label{eq:B_transf}
s_{x,\mu}u_{x,\mu}.
\end{equation}

At this point, the complete action in the 
coherent state path integral is
\begin{align}
S=S_B +\frac{Jn^4}{16} \sum_{x,\mu\neq \nu}&
2\text{Re }\text{Tr}\left[ s_{x,\mu}u_{x,\mu}
s_{x+\hat{\mu},\nu}u_{x+\hat{\mu},\nu}  
u^{\dag}_{x+\hat{\nu},\mu}s_{x+\hat{\nu},\mu}
u^{\dag}_{x,\nu}s_{x,\nu}\right] 
\notag \\
+&J'\sum_{x,\mu}\left[\det\left(\frac{n}{2}
s_{x,\mu}u_{x,\mu}\right)
+\det\left(\frac{n}{2}
s_{x,\mu}u^{\dag}_{x,\mu}\right)\right] 
\label{eq:gauge_ac}
\end{align}
Here, 
\begin{equation}
S_B=-\frac{n}{2}
\sum_{x,\mu} \int_0^{\beta} d\tau \,\text{Tr}
\left( \cos (2b_{x,\mu}) u_{x,\mu}\partial_{\tau}
u^{\dag}_{x,\mu}\right),
\end{equation}
which follows from (\ref{eq:berry1}).

We now want to expand around the minimum of the action. From 
(\ref{eq:gauge_ac}), the action is minimized when the eigenvalues of 
$s$ are largest, i.e.\ equal to one, and the field $u$ is constant for
all links on the lattice. We can use a gauge transformation to rotate
these constant fields to the identity matrix. The expansion for $s$
is the same as in section \ref{sec:contlim},
\begin{equation}
s_{x,\mu}= \sin (2b_{x,\mu}) 
 = \sin( 2(b_0 +\epsilon_{x,\mu})) 
 \approx \1 -2\epsilon^2_{x,\mu}.
\end{equation}

We substitute this expression into the action,
dropping terms of order $\epsilon^3_{x,\mu}$ and higher.
After some rearrangement and relabeling of
the summed indices, we find 
\begin{multline}
S\approx S_B 
+\frac{Jn^4}{16}\sum_{x,\mu\neq\nu}
2\text{Re }\text{Tr}\bigg[ \left(  \1 
-4 (\epsilon^2 _{x,\mu}+\epsilon^2_{x,\nu})
\right) 
u_{x,\mu}u_{x+\hat{\mu},\nu} 
 u^{\dag}_{x+\hat{\nu},\mu}
u^{\dag}_{x,\nu}\bigg] \\ 
+2J' \left(\frac{n}{2}\right)^{N}\sum_{x,\mu}
\left( 1 -2\,\text{Tr}\,\epsilon^2_{x,\mu}\right) 
\cos\theta_{x,\mu}.
\end{multline}
Here, we have defined $\exp (i\theta_{x,\mu})\equiv
\det u_{x,\mu}$. If we take $J'<0$, then the minimum
of the action will occur for $\theta_{x,\mu}=0$.
Thus, the matrices $u_{x,\mu}$ will have
determinant equal to one, and belong to
$SU(N)$ rather than $U(N)$. 

In order to be able to take the continuum limit we 
need to express the unitary matrix field $u_{x,\mu}$, assumed
to be close to the identity in our expansion, 
as the exponential of algebra-valued matrix fields.
So let
\begin{equation}
\aaa_{x,\mu}=-\aaa_{x,\mu}^b\lambda^b
\end{equation}
be a Lie algebra valued vector field, $\lambda^a$
denoting the generators of $SU(N)$, and write
\begin{equation}
\label{eq:unitary}
u_{x,\mu}\equiv \exp (-ia^2(\theta_{x,\mu}/N)\1
-ia\,\aaa_{x,\mu}).
\end{equation}
We also rescale the field $\epsilon_{x,\mu}$, writing
$\epsilon_{x,\mu}=a^2 \tilde{\epsilon}_{x,\mu}$.
The next step is to expand the action to order $a^4$, using
(\ref{eq:unitary}).
This leads to
\begin{equation}
S=S_B-\frac{Jn^4}{16}\sum_{x,\mu\neq\nu}a^4
\text{Tr}\left[ f_{\mu\nu}f_{\mu\nu}\right] 
+\sum_{x,\mu}a^4\text{Tr}\left[ \gamma\tilde{\epsilon}^2_{x,\mu}
-2J'\left(\frac{n}{2}\right)^N \theta^2_{x,\mu}\right],
\label{eq:S_expand}
\end{equation}
where we have dropped a constant, and
\begin{equation}
\gamma=-3Jn^4-4J'\left(\frac{n b}{2}\right)^N. 
\end{equation}
Of course, there aren't any terms linear in $\tilde{\epsilon}$ in 
(\ref{eq:S_expand}), and $\gamma>0$ since we are 
expanding about a minimum.

We use the same expansion to manipulate
the Berry phase term,
\begin{eqnarray}
S_B&=&-\frac{n}{2}\int_0^{\beta}d\tau\sum_{x,\mu}
\text{Tr}\left[
\cos(2b_{x,\mu})u_{x,\mu}\partial_{\tau}u^{\dag}_{x,\mu}
\right] \notag \\
&\approx&-\frac{n}{2}\int_0^{\beta}d\tau\sum_{x,\mu}
\text{Tr}\left[
\left( \cos(2b_0)-2\sin(2b_0)\epsilon_{x,\mu}\right) 
\right. \notag \\
 & &\left. \hspace{4cm}\times
(\1-ia\,\aaa_{x,\mu})\partial_{\tau}
(\1+ia\,\aaa_{x,\mu})
\right] \notag \\
&=&i\frac{n}{2}\int_0^{\beta}d\tau\sum_{x,\mu} a^4
\,\text{Tr}\left[ \frac{2}{a}\tilde{\epsilon}_{x,\mu}
\partial_{\tau}\aaa_{x,\mu} \right]. 
\end{eqnarray}

The complete action in the limit
$\sum_{x}a^4\to \int d^4 x$ is now
\begin{eqnarray}
S&=&\frac{1}{2e^2} \int_0^{\beta}d\tau 
\int d^4 x\, \text{Tr}
\bigg[ f_{\mu\nu}f_{\mu\nu}+2e^2\gamma\tilde{\epsilon}_{\mu}
\tilde{\epsilon}_{\mu}+2ie^2\frac{n}{a}
\tilde{\epsilon}_{\mu}\partial_{\tau}\aaa_{\mu} \notag \\
& &\hspace{6.75cm}
-4e^2 J' \left(\frac{n}{2}\right)^N \theta_{\mu}
\theta_{\mu}\bigg]
\notag \\
&=&\frac{1}{2e^2}\int_0^{\beta}d\tau 
\int d^4 x\, \text{Tr}
\bigg[ f_{\mu\nu}f_{\mu\nu} 
+\frac{1}{c^2}\partial_{\tau} \aaa_{\mu}
\partial_{\tau} \aaa_{\mu} 
+2e^2 \gamma \tilde{\epsilon}_{\mu}'
\tilde{\epsilon}_{\mu}' \notag \\
& &\hspace{6.75cm}
-4e^2 J'\left(\frac{n}{2}\right)^N 
\theta_{\mu}\theta_{\mu}
\bigg],
\end{eqnarray}
where $e^2=8/(n^4 |J|)$ and $c=(na/2)\sqrt{\gamma |J|}$.
We have completed the square in order to integrate 
out the shifted field
\begin{equation}
\tilde{\epsilon}_{\mu}'=\tilde{\epsilon}_{\mu}
+i\frac{n}{2\gamma a}\partial_{\tau}\aaa_{\mu},
\end{equation}
as well as the $\theta$-field, obtaining
\begin{equation}
S=\frac{1}{2e^2} 
\int_0^{\beta}d\tau \int d^4 x\, \text{Tr}
\left[ f_{\mu\nu}f_{\mu\nu}
+\frac{1}{c^2}\partial_{\tau} \aaa_{\mu}
\partial_{\tau} \aaa_{\mu}\right].
\end{equation}

If we now again assume that the correlation length is
much larger than the extent of the fifth dimension,
we can perform the trivial integration over
$\tau$, to obtain
\begin{equation}
S=\frac{\beta}{2e^2}\int d^4 x \,
\text{Tr}\left[f_{\mu\nu}f_{\mu\nu} \right].
\end{equation}

It was argued in \cite{ChWi97}, that a finite 
correlation length
\begin{equation}
\xi\propto \exp\left( \frac{24 \pi^2 \beta}{11Ne^2}
\right)
\end{equation}
is expected to be generated non-perturbatively.
Again, the continuum limit in which the 
correlation length diverges is achieved by taking
the extent of the extra dimension $\beta$
to infinity. In this limit, we also find that
the extent of the extra dimension in physical units
is much smaller than the correlation length,
$\beta c\ll \xi$. Thus, the theory undergoes 
dimensional reduction.

\section{Conclusions}
\label{ch:conclusions}
In the D-theory formulation of quantum field theories, a field
Lagrangian is replaced by a Hamilton operator and continuous
classical fields are replaced by operator fields. The Hamilton
operator evolves the system in an additional Euclidean direction.
Guided by symmetry considerations, we have formulated the 
principal chiral model as such a quantum spin system. 
We then went
on to show that with a particular choice of representation for the 
operators in the Hamiltonian, the theory reduces to a Wilsonian
lattice principal chiral model. From numerical simulations we know
that the $SU(N)_L\times SU(N)_R\times U(1)_{L=R}$ 
symmetry of such a model breaks spontaneously to
$SU(N)_{L=R}$ at $\beta=\infty$ \cite{Dreh92}. 
We chose representations with rectangular Young tableaux, with $N$
rows and $n$ columns, where $n$ was taken to be large.
The Goldstone modes arising from the spontaneous symmetry breaking
cause the system
to undergo dimensional reduction when we make the extent of the third
dimension finite and we thus recover the 2-d principal chiral model. 
We have seen that the
continuous degrees of freedom of the low-energy effective theory,
which is the same as the standard formulation of the principal
chiral model, 
arise as collective excitations of the discrete degrees of freedom
in the D-theory formulation of the model.

We also showed that the quantum link model in $(4+1)$-d undergoes
dimensional reduction to $4$-d Yang Mills theory. 
We chose the quantum link operators to be in the same type of representation
that we considered for the operators in the principal chiral model,
namely large representations with rectangular Young tableaux.
The mechanism for dimensional reduction is different in this
case. Instead of Goldstone modes arising from a spontaneously broken
global symmetry, the massless modes we need for dimensional reduction
result from the fact that a $(4+1)$-d gauge theory can exist in a
non-Abelian Coulomb phase. We showed that for the aforementioned 
representations the low-energy effective theory of the D-theory is 
indeed a 5-d Wilson-type
lattice gauge theory. It is known from numerical simulations
that such a theory is indeed in the non-Abelian Coulomb phase when the extent
of the fifth dimension is infinite \cite{BBCW98,Creu79}. At finite temperature 
the gauge bosons form glueballs and
acquire mass, due to the confinement hypothesis. 
The correlation length, however, is exponential in the extent of the 
fifth dimension,
hence leading to dimensional reduction. Again, the continuous fields of the
low-energy effective theory after dimensional reduction arise as collective 
excitations of discrete variables.

In order to be able to get an analytic handle on the behavior of the D-theory
formulations of the principal chiral model and non-Abelian gauge theory,
we had to consider large representations for the quantum operators
in the Hamiltonian. 
On the other hand, to develop more efficient algorithms for simulating
such theories one would like to consider smaller representations, so that
each variable can assume only a few discrete values. It is not clear at this
point if the mechanism of dimensional reduction also occurs for small
representations. Numerical studies are needed to answer this
question.

\begin{appendix}
\section{Appendix}
For completeness' sake, we show in this section how to 
decompose a non-singular $GL(N,\mathbb{C})$ matrix $q$
into the product of a Hermitian matrix $b$ and a unitary
matrix $u$.
First, let $m=qq^{\dag}$, which
is Hermitian positive semidefinite. So it can be
diagonalized by a unitary transformation,
\begin{equation}
vmv^{\dag}=m_D=\mbox{diag}(m_1,m_2,\ldots,m_N).
\end{equation}
We can then define the square root of $m_D$ as
\begin{equation}
\sqrt{m_D}=\mbox{diag}(\sqrt{m_1},\sqrt{m_2},\ldots,
\sqrt{m_N}).
\end{equation}
Now let
\begin{equation}
b= v^{\dag}m_D^{1/2}v, \hspace{0.5cm}
u= b^{-1}q=v^{\dag}m_D^{-1/2}vq.
\end{equation}
It is easily seen that such a $u$ is unitary. We also want
to determine the transformation properties of the $b$ and
$u$ matrices if $q$ transforms under $U(N)_L\times
U(N)_R$ transformations as 
\begin{equation}
q\to q'=LqR^{\dag},
\end{equation}
where $LL^{\dag}=RR^{\dag}=\1$. 
Then $q'=b'u'$, and 
\begin{equation}
m'=q'q^{\prime\dag}=LqR^{\dag}Rq^{\dag}L^{\dag}=
LmL^{\dag}.
\end{equation}
We also have,
\begin{equation}
m_D'=v' m' v^{\prime\dag}=v'LmL^{\dag}v^{\prime\dag}=m_D.
\end{equation}
Thus, $v'L=dv$ or $v'=dvL^{\dag}$, where $d$ is a 
non-degenerate diagonal matrix. So we find the
transformation properties
\begin{equation}
b'=v^{\prime\dag}\sqrt{m_D}v'=Lv^{\dag}d^{\dag}
\sqrt{m_D}dvL^{\dag}=LbL^{\dag}.
\end{equation}
and
\begin{equation}
u'=b^{\prime-1}q'=Lb^{-1}L^{\dag}LqR^{\dag}=LuR^{\dag}.
\end{equation}
Furthermore, observe that 
\begin{equation}
qq^{\dag}=m=v^{\dag}m_D v=(v^{\dag}\sqrt{m_D}v)
(v^{\dag}\sqrt{m_D}v)=b^2, \hspace{0.5cm}
q^{\dag}q= u^{\dag} b^2 u.
\end{equation}
\end{appendix}



\begin{thebibliography}{10}

\bibitem{ChWi97} S.~Chandrasekharan and U.-J.~Wiese, 
                 Nucl.\ Phys.\ B492 (1997) 455. 

\bibitem{BBCW98} B.~B.~Beard, R.~C.~Brower, S.~Chandrasekharan, D.~Chen, 
                 A.~Tsapalis, \mbox{U.-J.~Wiese}, Nucl.\ Phys.\ Proc.\ Suppl.\ 
                 63 (1998) 775.

\bibitem{BChW97} R.~Brower, S.~Chandrasekharan and 
                 U.-J.~Wiese, Phys.\ Rev.\ D60 (1999) 094502.

\bibitem{Wies98} U.-J.~Wiese, Nucl.\ Phys.\ Proc.\ Suppl.\ 73
                 (1999) 146.

\bibitem{MeWa66} N.~D.~Mermin and H.~Wagner, Phys.\ Rev.\ 
                 Lett.\ 17 (1966) 1133; \\
	         P.~C.~Hohenberg, Phys.\ Rev.\ 158 (1967) 383; \\
                 S.~Coleman, Comm.\ Math.\ Phys.\ 31 (1973)
                 259.

\bibitem{CHaN88} S.~Chakravarty, B.~I.~Halperin and D.~R.~Nelson, 
                 Phys.\ Rev.\ Lett. 60 (1988) 1057; 
                 Phys.\ Rev.\ B39 (1989) 2344.

\bibitem{Has91}  P.~Hasenfratz and F.~Niedermayer, 
                 Phys.\ Lett.\ B268 (1991) 231.

\bibitem{KoCh89} P.~Kopietz and S.~Chakravarty, Phys.\ Rev.\ B40
                 (1989) 4858.

\bibitem{ScWi00} B.~Schlittgen and U.-J.~Wiese, Nucl.\ Phys.\ 
                 Proc.\ Suppl.\ 83-84 (2000) 718.

\bibitem{Creu79} M.~Creutz, Phys.\ Rev.\ Lett.\ 43 (1979) 553.

\bibitem{Sham93} Y.~Shamir, Nucl.\ Phys.\ B406 (1993) 90.

\bibitem{FuSh95} V.~Furman and Y.~Shamir, Nucl.\ Phys.\ B439
                 (1995) 54.

\bibitem{Kapl92} D.~B.~Kaplan, Phys.\ Lett.\ B288 (1992) 342.

\bibitem{ReSa89} N.~Read and S.~Sachdev, Nucl.\ Phys.\ 
                 B316 (1989) 609.

\bibitem{DaSS93} S.~Randjbar-Daemi, A.~Salam and J.~Strathdee, Phys.\ Rev.\
                 B48 (1993) 3190.

\bibitem{DaSt93} S.~Randjbar-Daemi and J.~Strathdee, Int.\ J.\ Mod.\ Phys.\
                 A8 (1993) 3509.

\bibitem{Horn81} D.~Horn, Phys.\ Lett.\ 100B (1981) 149.

\bibitem{OrRo90} P.~Orland and D.~Rohrlich, Nucl.\ Phys.\ B338 (1990) 647.

\bibitem{Chan99} S.~Chandrasekharan, Nucl.\ Phys.\ Proc.\ Suppl.\ 73 
                 (1999) 739.

\bibitem{Barn91} T.~Barnes, Int.\ J.\ Mod.\ Phys.\ C2 (1991) 659.

\bibitem{WiYi94} U.-J.~Wiese and H.-P.~Ying, Z.\ Phys.\ B93 (1994) 147.

\bibitem{Dreh92} P.~Dreher, Phys.\ Lett.\ B281 (1992) 127;
                 Nucl.\ Phys.\ B389 (1993) 265.

\bibitem{Auer94} A.~Auerbach, {\em Interacting
                 Electrons and Quantum Magnetism}
                 (Springer, New York, 1994).

\bibitem{Leut94} H.~Leutwyler, Phys.\ Rev.\ D49 (1994) 3033.

\bibitem{Pere86} A.~Perelomov, {\em Generalized 
                 Coherent States
                 and Their Applications} (Springer, 
                 Berlin, 1986).

\bibitem{Feyn51} R.~P.~Feynman, Phys.\ Rev.\ 84 (1951) 108.

\bibitem{Wieg84} P.~B.~Wiegmann, Phys.\ Lett.\ 
                 141B (1984) 217.

\bibitem{Wils74} K.~Wilson, Phys.\ Rev.\ D10 (1974)
                 2445.

\end{thebibliography}
\end{document}